\documentclass{PoS}

\newcommand{\bce}{\begin{center}} 
\newcommand{\ece}{\end{center}}
\newcommand{\beq}{\begin{equation}}
\newcommand{\eeq}{\end{equation}}
\newcommand{\bea}{\vspace{0.25cm}\begin{eqnarray}}
\newcommand{\eea}{\end{eqnarray}}

\newcommand{\brho}{\mbox{\boldmath $\rho$}}

\newcommand{\br}{{\bf r}}

\newcommand{\do}{\!\downarrow}

\newcommand{\ba}{\begin{array}}
\newcommand{\ea}{\end{array}}

\newcommand{\bb}{{\bf b}}
\def\lsim{\mathrel{\rlap{\lower4pt\hbox{\hskip1pt$\sim$}}
    \raise1pt\hbox{$<$}}}         
\def\gsim{\mathrel{\rlap{\lower4pt\hbox{\hskip1pt$\sim$}}
    \raise1pt\hbox{$>$}}}         

\def\beq{\begin{equation}}
\def\endeq{\end{equation}}
\def\arr{\begin{eqnarray}}
\def\endarr{\end{eqnarray}}

\title{Left and Right in small-x neutrino DIS}

\ShortTitle{Left and Right in small-x neutrino DIS}

\author{\speaker{Vladimir Zoller}\\%
        ITEP, Moscow\\
        E-mail: \email{zoller@itep.ru}}

\abstract{The color dipole analysis of nuclear 
effects in charge current DIS is presented. The emphasis is  put on
the pronounced effect of  left-right  asymmetry of shadowing in
neutrino-nucleus DIS at small values of Bjorken $x$. 
 Strikingly
different scaling behavior of nuclear shadowing for the left-handed and
right-handed $W^+$ is  predicted. Large, about $25\%$, shadowing in the 
$Fe$ structure functions
is predicted, which is important for a precise determination of
the P-odd nucleon  structure functions $xF^{\nu(\bar\nu)}_3$ and 
$\Delta x F_3=xF_3^{\nu}-xF_3^{\bar \nu}$.}

\FullConference{Diffraction 06, International Workshop on 
Diffraction in High-Energy Physics\\
                 September 5-10, 2006\\
                 Adamantas, Milos island, Greece}

\begin{document}
\section{Introduction}
Responsible for the account of the   left-right asymmetry effects  in 
diffractive  neutrino  interactions with 
nuclear medium is the parity-odd structure function $F_3$ which appears  in
the  standard decomposition of  the forward amplitude for virtual
W-nucleon scattering \cite{Ioffe}
\beq
{1\over \pi}W_{\mu\nu}=...+i\varepsilon_{\mu\nu\lambda\rho}
{p_\lambda q_\rho\over 2pq}F_3(x,Q^2)...,
\label{eq:Compton}
\eeq
where $p$ and $q$ are the target nucleon and  the virtual 
$W^+$-boson four-momenta, respectively. 
 The physical picture behind the decomposition (\ref{eq:Compton}) 
depends on the choice of particular reference frame.  
Indeed, in the brick wall frame, where the parton picture of a
 nucleon is manifest, in the chiral 
limit the left-handed W-boson ($W_L$) interacts only with quarks 
while the right-handed $W_R$ interacts only 
with anti-quarks. Hence, the representation for $F_3$ in terms of small-x parton 
densities \cite{Ioffe}
\bea
F_3^{\nu N}=2(s-{\bar c}) \\
F_3^{{\bar\nu} N}=2(c-{\bar s})\,.
\label{eq:F3ISO}
\eea
 Thus, the isospin symmetry of the nucleon sea 
implies that at small Bjorken $x$ 
the  structure function $F_3$ is dominated by 
the charm-strange   weak current.

In the dipole/laboratory  frame
 the
  small-$x$ DIS is treated in terms of the
interaction of the quark-antiquark color dipole (CD) of size ${\bf r}$,
that 
the virtual $W$ transforms into,  
 with the target nucleon \cite{NZ91,M}.  This interaction is described by the universal 
 flavor-independent color dipole cross section
$\sigma(x,r)$. At small $x$ the dipole size ${\bf r}$ 
 is a conserved quantum number and the parity non-conservation effect can be 
quantified in terms of CD sizes of left-handed and right-handed 
W-bosons. Hence, the representation of $F_3$ in terms of
absorption cross sections for $W_L$- and $W_R$-bosons ($\sigma_{L,R}$)
\beq
2xF_3(x,Q^2)={Q^2\over 4\pi^2\alpha_W}\left(\sigma_L-\sigma_R\right)\,,
\label{eq:F3SIGMA}
\eeq 
where $\alpha_W=G_Fm^2_W/4\pi\sqrt{2}$.
These cross sections
 are calculated as a quantum mechanical expectation values of
$\sigma(x,r)$. For small dipoles 
\beq
\sigma(x,r)\propto  r^2\alpha_S(r^2)G(x,C/r^2)
\label{eq:Small}
\eeq 
where $C\simeq 10$ and  $G(x,\kappa^2)$ is 
the gluon density in the proton. 
Then, the deviation of $F_3$ from zero implies that dipole sizes of 
left-handed and right-handed W-bosons are different. 
The origin of this difference is of course  the left-handedness of 
weak currents. The latter makes  the   light-cone densities of 
charm-strange dipoles highly asymmetric functions of the Sudakov variable
$z$ \cite{FZ1,FZ2},
\beq
|\Psi_L(z,r)|^2\sim z^2\varepsilon^2 K^2_1(\varepsilon r)
\label{eq:PsiL}
\eeq 
 and 
\beq
|\Psi_R(z,r)|^2\sim (1-z)^2\varepsilon^2 K^2_1(\varepsilon r)
 \label{eq:PsiR}
\eeq   
 Hereafter $z$ stands for 
 the momentum fraction of the light-cone $W$-boson carried away by
the $c$ quark. 
Characteristic values of $z$ for left-handed and right-handed
 $c\bar s$ states are very different. Besides,  the dipole sizes are limited 
from above 
by the factor  $K^2_1(\varepsilon r)\sim \exp(-2\varepsilon r)$ and 
the cutoff parameter $\varepsilon$ depends on $z$,
\beq
\varepsilon^2=z(1-z)Q^2+(1-z)m^2_c+zm^2_s.
\label{eq:epsil}
\eeq   
Therefore,  different $z$ correspond to  different
dipole sizes dominating  the integrals  for
the dipole-nucleon cross sections  $\sigma_L$ and $\sigma_R$ at $Q^2\gg m_c^2$,
\bea
\sigma_{L}(x,Q^{2})
=\int_0^1 dz \int d^{2}{\bf{r}} 
|\Psi_{L}(z,{\bf{r}})|^{2} 
\sigma(x,r)  \nonumber\\
\propto {1\over Q^{2}}\int_{1/Q^2}^{1/m_s^2} {dr^{2}
\over r^4}\sigma(x,r)\nonumber\\
\propto {1\over Q^{2}}\int_{1/Q^2}^{1/m_s^2} {dr^{2}\over r^2}
\alpha_S(r^2)G(x,10/r^2)\propto {1\over Q^2}\log{Q^2\over m_s^2}
\label{eq:SIGMAL}
\eea
\bea
\sigma_{R}(x,Q^{2})
\propto {1\over Q^{2}}\int_{1\over Q^2}^{1/m_c^2} {dr^{2}
\over r^4}\sigma(x,r)\nonumber\\
\propto {1\over Q^{2}}\int_{1/Q^2}^{1/m_c^2} {dr^{2}\over r^2}
\alpha_S(r^2)G(x,10/r^2)\propto {1\over Q^2}\log{Q^2\over m_c^2}.
\label{eq:SIGMAR}
\eea
where $m_c$ and $m_s$ are constituent quark masses. In Eqs.(\ref{eq:SIGMAL},
\ref{eq:SIGMAR}) it has been taken
 into account that the product
$\alpha_S(r^2)G(x,C/r^2)$ is approximately  flat in $r^2$.
 
The mirror asymmetry of  diffractive
interactions of electroweak bosons with  nuclear matter
 is enhanced
by the large thickness of a nucleus \cite{FZ2,ichep33}.  In particular, 
the nuclear shadowing effect appears  to be 
very different for left-handed and 
right-handed $W$-bosons. The point is that the shadowing term 
$\delta\sigma_{A,\lambda}$, where  $\lambda=L,R=-1,+1$,
 in expansion of the nuclear cross section
\beq
\sigma_{A,\lambda}=
A\sigma_{\lambda}
-\delta\sigma_{A,\lambda}.
\label{eq:NUCLEAR}
\eeq
is the quadratic functional of the CD  cross section. 
To the lowest order  in $\sigma T$ this  term  reads 
\beq
\delta\sigma_{A,\lambda}\simeq
{\pi\over 4} \langle \sigma^2\rangle_{\lambda}
\int db^2 T^2(b),
\label{eq:SHADOW}
\eeq
Here 
$T(\bb)=\int_{-\infty}^{+\infty} dz n_A(\sqrt{z^2+\bb^2}),
$
is the optical  thickness of a nucleus  
at an impact parameter $\bb$, the nuclear matter density $n_A(\br)$
is normalized as  
$\int d^3\br n(\br)=A $.
The leading-log contribution to $\langle \sigma^2\rangle_{\lambda}$ comes from
the domain 
\beq
{1\over Q^2}\lsim r^2\lsim r_\lambda^2,
\label{eq:Domain}
\eeq
where $r_L^2=1/m_s^2$  and  $r_R^2=1/m_c^2$. Then,
\bea
\langle \sigma^2\rangle_{L}=
\langle\Psi_{L}|\sigma(x,r)^2|\Psi_{L} \rangle
=\int dz d^{2}{\bf{r}} 
|\Psi_{L}(z,{\bf{r}})|^{2} 
\sigma^2(x,r)\nonumber\\
\propto  {1\over Q^2}\int_{1/Q^2}^{1/m_s^2} {dr^{2}
\over r^4}\sigma^2(x,r)\propto {1\over {Q^2m_s^2}}
\label{eq:SIG2L}
\eea  
and, similarly,
\bea
\langle \sigma^2\rangle_{R}
\propto  {1\over Q^2}\int_{1/Q^2}^{1/m_c^2} {dr^{2}
\over r^4}\sigma^2(x,r)\propto {1\over {Q^2m_c^2}}.
\label{eq:SIG2R}
\eea 
 Thus, there is a sort of filtering phenomenon,
the target nucleus absorbs the $c\bar s$ Fock component of $W^+$
 with the helicity $\lambda=-1$, but 
is nearly transparent for $c\bar s$ states with opposite helicity 
\cite{FZ2,ichep33}.

\section{Color dipole description of CC DIS off nuclei: 
first iteration of the LL$(1/x)$ evolution \cite{Log1/x}}

We base our calculations of nuclear cross sections
 on the CD  approach to the Leading 
Log$(1/x)$ (LL$(1/x)$), 
or BFKL \cite{BFKL}, evolution of DIS \cite{NZZBFKL,NZ94}.
For nuclear targets a complete resummation of LL$(1/x)$ 
effects is as yet lacking.
 Our point is that at energies
of the  planned
eIC \cite{eRHIC}, the $x$-dependence of nuclear shadowing  (NS)
is practically exhausted by the 
first CD LL$(1/x)$ iteration which is calculable exactly
without invoking the large-$N_c$ 
approximation. Indeed,  the 
first CD LL$(1/x)$ iteration dominates in 
the $x$ region
\beq
\xi = \log {x_0 \over x } \lsim {1\over \Delta_{eff}}, 
\label{eq:REGION} 
\eeq
where $\Delta_{eff}\simeq 0.1-0.2$ is the exponent of the
local $x$-dependence of the proton structure function,
$F_{2p} \propto x^{-\Delta_{eff}}$ \cite{HERAdeltaPom}. 

When viewed in the laboratory frame, NS corrections to $F_3$ come 
 from the coherent 
interaction of $c\bar s, c\bar s g,...$
states. To the required accuracy, the Fock state expansion 
of the virtual W boson  $|W^*\rangle$ reads
\bea
|W^*\rangle=\sqrt{Z_g} \Psi_{\lambda}|c\bar s\rangle+
\Phi_{\lambda}|c\bar s g\rangle,
\label{eq:FOCK}
\eea
where $\Psi_{\lambda}$ and 
$\Phi_{\lambda}$ 
are the light-cone wave functions (WF's)
of the $c\bar s$ and $c\bar s g$ states \cite{FZ1,FZ2},
$\sqrt{Z_g}$ is the renormalization of the  
$c\bar s$ state by the virtual radiative corrections 
for the $c\bar s g$ state. For soft gluons
the 3-parton WF takes the 
factorized form 
$\Phi_{\lambda}=\Psi_{\lambda}\{\Psi_{cg}-\Psi_{{\bar s}g}\}$ 
\cite{NZZBFKL,NZ94}. 
The nuclear coherency condition reads
\beq
{x\over \beta} \lsim x_A = {1\over {m_N R_A}} =0.15 A^{-1/3},
\label{eq:COHERENCY}
\eeq
where  $x=Q^2/2m_N\nu$ is the Bjorken variable, 
$\nu$ is the energy of the photon, 
$M$ is the invariant mass of the multiparton Fock state
and $\beta= Q^2/(M^2 + Q^2)$. 

In the CD
BFKL-Regge phenomenology of DIS one usually formulates the boundary 
condition 
at $x= x_0=0.03$ \cite{BFKLRegge}. 
For extremely heavy nuclei
$x_A\ll x_0$
and the LL$(1/x)$ evolution starts at $x=x_A$ 
with the boundary condition formulated in terms of the free-nucleon
quantities  LL$(1/x)$-evolved from $x_0$ down to $x_A$. 
At small $x$ satisfying the condition (\ref{eq:COHERENCY}) the 
color dipoles $\{\br_n\}$ in the multiparton state 
are conserved in the interaction process.  At the boundary $x=x_A$ and for
$A\gg 1$, the nuclear 
$\textsf{S}$-matrices equal \cite{Glauber,Gribov} 
\beq 
\textsf{S}_{n}(x_A,\bb,\{\br\}_n) = \exp[-{1\over 2} \sigma_n(x_A,\{\br\}_n)T(\bb)],
\label{eq:SMATRIXC}
\eeq 

\begin{figure}
\includegraphics[width=0.9\textwidth]{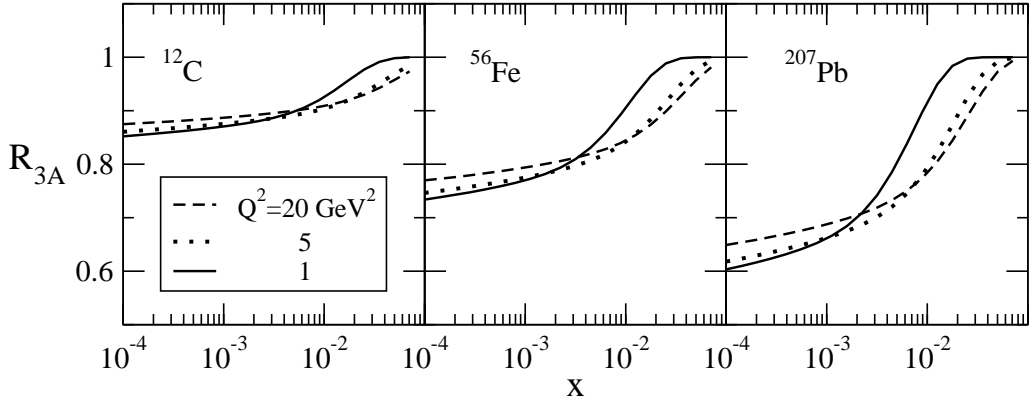}
\caption{Predictions from CD LL$(1/x)$ evolution
for NS ratio $R_{3A}$ for C, Fe and Pb nuclei as a function of $x$ at  
$Q^2=1,5,20$ GeV$^2$.}
\label{fig:RNUC}
\end{figure} 

where  $\sigma_n(x_A,\{\br\}_n)$ is the free-nucleon CD cross section for the
$n$-parton state, and \cite{NZ9193}
\beq
\sigma_{2,A}(x_A,\br )=2\int d^2\bb[1 - \textsf{S}_{2}(x_A,\bb,\br)].
\label{eq:SIGMA0}
\eeq

Now  one 
must evolve the nuclear $\textsf{S}$-matrix down to $x\ll x_A$. Specifically,
after the gluon variables have been properly integrated out, the effect of the
extra gluon in the Fock state expansion for the incident photon boils down
to the renormalization of the $\textsf{S}$-matrix and nuclear cross section 
for the $c\bar{s}$ CD \cite{NZZBFKL,NZ94},:
\bea
{\partial{\textsf S}_2(x,\br;\bb) \over \partial \xi }
&=&\int d^2\brho  |\psi(\brho) -\psi(\brho+\br)|^2  \Big[
{\textsf S}_{3}(x,\brho,\br;\bb) - {\textsf S}_{2}(x,\br;\bb) \Big], \nonumber\\
\sigma_A(x,\br) &=& \sigma_{2,A}(x_A,\br ) +\sigma^{(1)}_A(x,\br ), \nonumber\\
\sigma^{(1)}_A(x,\br)& =& 2\log\left({x_A\over x}\right)
\int d^2\bb\int d^2\brho  |\psi(\brho) -\psi(\brho+\br)|^2\nonumber\\
&\times& \Big[ {\textsf S}_{2}(x_A,\br;\bb)- {\textsf S}_{3}(x_A,\brho,\br;\bb)\Big],\nonumber\\
\sigma_3(x,\br,\brho) &=& {N_c^2 \over N_c^2-1}\Big[\sigma_2(x,\brho) +
\sigma_2(x,\brho + \br)\Big] - {1 \over N_c^2-1} \sigma_2(x,\br)\label{eq:DGDXI}
\eea
where $\rho$ is the size of the  $cg$ dipole and 
\beq
\psi(\brho)={\sqrt{C_F\alpha_S}\over \pi}\cdot {\brho\over \rho^2 }\cdot {\rho
  \over R_c} K_1({\rho\over R_c})
\label{eq:PSI}
\eeq
is the radial WF of the $cg$ state with the Debye screening
of infrared gluons \cite{NZZBFKL,NZ94}.
The absorption  cross section is 
an expectation value 
\beq
\sigma_{A,\lambda}(x,Q^2) =
\langle \Psi_\lambda|\sigma_A(x,\br)|\Psi_\lambda
\rangle\,.  
\label{eq:SigN} 
\eeq
The shadowing ratio $R_{3A}$, equals  
\bea
R_{3A}={\sigma_{A,L}-\sigma_{A,R} \over  
A(\sigma_L- \sigma_R)} = 1- {\delta\sigma_{A,L}-\delta\sigma_{A,R}\over
A\sigma_L-A\sigma_R},
\label{eq:Rshad}
\eea 
where  $\sigma_{\lambda}(x,Q^2)=\sigma_{2,\lambda}(x_A,Q^2)+
\log(x_A/x)\sigma^{(1)}_{\lambda}(x_A,Q^2)$ is 
the LL$(1/x)$-evolved free-nucleon cross section. 
 Hence, the nuclear shadowing correction  
$(1-R_{3A})\cdot xF^{\nu(\bar \nu)}_3$
which should be added to  $xF^{\nu(\bar\nu)}_3$ extracted from
the $\nu(\bar\nu) Fe$ data to get the  ``genuine'' $ xF^{\nu(\bar\nu)}_3$. 
This correction positive-valued  and does
increase $xF^{\nu(\bar\nu)} _3$ of the impulse approximation.

Our interest is in the well 
evolved shadowing at $x\ll x_A$, the onset of NS at $x \gsim x_A$ must be treated
within the light-cone Green function technique \cite{BGZ98,BGZ96}, here
we show only the gross features of the large-$x$ suppression of NS
following the prescriptions from Refs. \cite{NZ9193,Karmanov} and the 
$\beta$-dependence of the large rapidity gap  DIS as predicted in \cite{NZ92} and
confirmed in the HERA experiments \cite{ZEUSdiff}. We use the free-nucleon
CD 
cross section tested against the
experimental data from HERA, it is described in  \cite{Log1/x}.
We calculate $R_{3A}$ as a function of $x$ for several values 
of  $Q^2$.
Our results obtained for  realistic nuclear densities of 
Ref.\cite{DEVRIES} are presented 
 in Figure \ref{fig:RNUC}.  Shown is  
the ratio $R_{3A}$ for different nuclear 
targets including $^{56}Fe$. 
At small $x$ and  high $Q^2$  the shadowing correction
scales, $\delta\sigma_{L,R}\propto 1/Q^2$.
 The absorption cross section $\sigma_{L,R}$ 
scales as well. The ratio $\delta\sigma_{L,R}/\sigma_{L,R}$ 
slowly decreases with growing $Q^2$ because 
of the logarithmic scaling violation in $\sigma_{L,R}$.
Toward the region of $x>0.01$,
 both the nuclear form factor and the mass threshold effect
suppress the shadowing correction $1-R_{3A}$ at $Q^2\lsim (m_c+m_s)^2$ 
(see Fig.\ref{fig:RNUC}).

\section{Summary}
The parity non-conservation effect in diffractive  charged current DIS was 
quantified in terms of color dipole sizes of left-handed and right-handed 
electroweak bosons.
We identified the origin  and estimated the strength of the left-right 
asymmetry effect. Based on the color dipole approach which correctly 
reproduces the experimental
data on the proton structure function measured at HERA and
the NMC data on nuclear shadowing,
we reported the quantitative predictions for the  LL$(1/x)$ 
evolution of the nuclear shadowing effect
 in absorption 
 of left-handed  and right-handed 
$W$-bosons.

\vspace{0.2cm} \noindent \underline{\bf Acknowledgments:}

The work has been partly supported by the DFG grant 436 RUS 17/82/06 and by 
 the RFBR grant 06-02-16905-a.

\end{document}